\documentclass[10pt,conference]{IEEEtran}
\usepackage{amsthm,amsmath}
\IEEEoverridecommandlockouts
\usepackage{graphicx, epsfig, epsf}
\usepackage[dvips]{color}   
\usepackage{mathrsfs}
\usepackage{amssymb}
\DeclareMathAlphabet{\mathbit}{OT1}{cmr}{bx}{it}

\newtheorem{cor}{Corollary}
\newtheorem{thm}{Theorem}

\renewcommand{\baselinestretch}{0.98}

\begin{document}

\title{Trapping Set Enumerators for Repeat Multiple Accumulate Code Ensembles}

\author{\IEEEauthorblockN{Christian Koller\IEEEauthorrefmark{1},
Alexandre Graell i Amat\IEEEauthorrefmark{2},
J{\"o}rg Kliewer\IEEEauthorrefmark{3}, and
Daniel~J.~Costello, Jr.\IEEEauthorrefmark{1}}\\

 \IEEEauthorblockA{
 \IEEEauthorrefmark{1}Department of
Electrical Engineering,
University of Notre Dame,
Notre Dame, IN 46556, USA\\
Email:  \{dcostel1, ckoller\}@nd.edu  }
\IEEEauthorblockA{
\IEEEauthorrefmark{2} Department of Electronics,
Institut TELECOM-TELECOM Bretagne,
29238 Brest, France\\
Email: alexandre.graell@telecom-bretagne.eu  }
\IEEEauthorblockA{\IEEEauthorrefmark{3}Klipsch School of Electrical and
  Computer Engineering,
  New Mexico State University,
  Las Cruces, NM 88003, USA\\
  Email:  jkliewer@nmsu.edu}

\thanks{This work was partly supported by NSF grants \mbox{CCF05-15012},
\mbox{CCF08-30666}, NASA grant NNX07AK536, and the Marie Curie Intra-European
Fellowship within the 6th European Community Framework Programme.} }

\maketitle

\begin{abstract}
The serial concatenation of a repetition code with two or more accumulators
has the advantage of a simple encoder structure. Furthermore, 
the resulting ensemble is asymptotically good and exhibits minimum distance
growing linearly with block length. However, in practice these codes cannot 
be decoded by a maximum likelihood decoder, and iterative decoding 
schemes must be employed. For low-density parity-check codes, the notion of
trapping sets has been introduced to estimate the performance of these codes
under iterative message passing decoding. 
In this paper, we present a closed form finite 
length ensemble trapping set enumerator for repeat multiple accumulate
codes by creating a trellis representation of trapping sets.
We also obtain the asymptotic expressions when the block length tends
to infinity and evaluate them numerically.

\end{abstract}

\section{Introduction}

Turbo-like codes \cite{ber93}, as well as LDPC codes \cite{Gal63}, can 
perform close to the Shannon limit using suboptimal iterative decoding 
schemes. However, these codes typically exhibit an error floor at medium 
to high signal-to-noise ratios (SNRs). In \cite{mac03}, 
the height of the error floor of LDPC codes was linked to 
so-called "near codewords". Later, in \cite{ric03}, this concept was
generalized to trapping sets, substructures in the Tanner 
graph of a code that may cause the iterative message passing decoder to fail. 
For certain LDPC codes, small trapping sets, rather than the minimum 
distance of the code, dominate the error floor performance.

Asymptotic spectra of trapping sets in LDPC code ensembles were 
computed in \cite{Mil07} for regular and irregular
LDPC codes and in \cite{AsRD07} for protograph-based codes. 
It was shown that there exist LDPC codes that 
exhibit a minimum trapping set size growing linearly with block length, 
for certain types of trapping sets.

In turbo-like codes, the concatenation of simple component codes through 
interleavers can lead to powerful code constructions. The simplest examples 
are repeat multiple accumulate (RMA) codes. These codes have a low encoding
complexity of $O(1)$ and can be decoded using relatively few iterations.
Furthermore, it has been shown in \cite{PfThs03} and \cite{all07} that 
the double serially concatenated repeat accumulate accumulate (RAA) 
code of rate 1/3 or smaller is asymptotically good and exhibits minimum 
distance growing linearly with block length.

Like LDPC codes, turbo-like codes are decoded in an iterative fashion.
Commonly, the component codes are decoded with a maximum a 
posteriori probability (MAP) decoding algorithm and the extrinsic information provided
by a component decoder functions as a priori information for another.
For RMA codes, the turbo decoder can be represented as a message passing
decoder \cite{KFL01}, similar to the belief propagation decoder, albeit 
with a different message passing schedule.
Thus the turbo decoder may also be susceptible to trapping sets.
To predict the error floor of a code one generally needs to have full knowledge of
the trapping sets that dominate the error floor, i.e., one needs to know their 
graph structure and enumerate their multiplicities, and to find the 
probability that the decoder gets trapped in a particular set. The latter 
not only depends on the graph structure of the trapping set but also on the 
channel model, the decoding algorithm, and the particular decoder implementation
that is used.

In this paper we address the first part of the problem, the enumeration of 
subgraphs in an RAA code. We derive a closed form trapping set enumerator (TSE) for 
general $(a,b)$ trapping sets, as defined in \cite{ric03} and \cite{Mil07}.
A general $(a,b)$ trapping set for a given Tanner graph is a set of $a$ variable nodes
that induces a subgraph containing $b$ odd degree check nodes,
which can be thought of as \textit{unsatisfied} checks,
and an arbitrary number of even degree check nodes. If there are
only a few unsatisfied check nodes and a sufficiently
large number of erroneous variable nodes, the iterative message passing decoder
may not be able to correct the erroneous nodes.
The TSE is the average number of $(a,b)$ trapping sets in the ensemble 
composed of all possible interleaver realizations. 
We also derive asymptotic expressions for the TSE and analyze them.

\section{Trapping set enumerators for repeat accumulate accumulate code ensembles}
\label{sec:trapenum}

\begin{figure}[t]
   \centering
    \includegraphics[width=0.99\columnwidth]{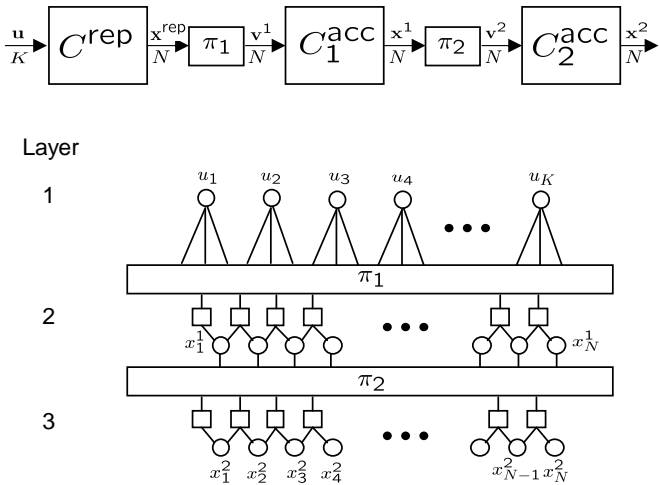}
    \vspace{-2ex}
    \caption{Block diagram and factor graph ($q=3$) of an RAA encoder.
    				 The circles represent variable nodes and the boxes
						 check nodes, respectively. }
    \label{fig:RAAEncoder}
    \vspace{-2ex}
\end{figure}

The encoder structure of an RAA code $\mathcal{C}^\mathrm{RAA}$ 
is shown in Fig.~\ref{fig:RAAEncoder}. It is a
serial concatenation of a repetition code $\mathcal{C}^\mathrm{rep}$ of rate
$R_\mathrm{rep}=1/q$ and two identical rate-1, memory-1, accumulate
codes $\mathcal{C}^{\mathrm{acc}}_l$, $l=\{1,2\}$, with generator polynomials
$g(D)=1/(1+D)$, connected by interleavers $\pi_1$ and $\pi_2$. 
For the repetition code $\mathcal{C}^\mathrm{rep}$, we denote
the binary input sequence of length $K$ by $\mathbf{u}=[u_1,\ldots,u_K]$
and the binary output sequence of length $N$ by
$\mathbf{x}^\mathrm{rep}=[x_1^{\mathrm{rep}},\ldots,x_N^{\mathrm{rep}}]$.
Likewise, for encoder $\mathcal{C}^{\mathrm{acc}}_l$,
$\mathbf{v}^l=[v^l_1,\ldots,v^l_N]$ and
$\mathbf{x}^l=[x^l_1,\ldots,x^l_N]$ denote the input sequence and
the codeword, respectively, where both are of length $N$.
Note that
$\mathbf{v}^1=\pi_1(\mathbf{x}^\mathrm{rep})$ and
$\mathbf{v}^2=\pi_2(\mathbf{x}^1)$. The overall code rate is
$R=K/N$.

The factor graph of an RAA code
is also depicted in Fig.~\ref{fig:RAAEncoder} for a repetition 
factor of $q=3$. The circles represent variable
nodes while the boxes represent check nodes. The information
symbols $\mathbf{u}$ correspond to the variable nodes of the first
layer and their degree is equal to the repetition factor $q$.
The variable nodes of the second layer correspond to the output
of the first accumulator $\mathbf{x}^1$, and the variable nodes
of the third layer to the output of the second accumulator
$\mathbf{x}^2$, respectively.
The input bits of the accumulators are represented by the variable 
nodes of the next higher layer. Only the variable nodes in 
the third layer are transmitted through the channel.
This implies that, initially, only variable nodes in the third layer 
can be in error, while the others have a neutral initial value.
However, in the first decoding iteration, the variable nodes in 
layers 1 and 2 get values assigned based on the received sequence.
If there are trapping sets containing variable nodes in those layers,
erroneous values that were assigned during the first iteration may never
be corrected and may cause the iterative decoder to fail.
Therefore, we consider the whole graph when enumerating for trapping sets.

Let $\bar{A}^{\mathcal{C}^\mathrm{RAA}}_{a,b}$ be the
ensemble-average TSE of an RAA code
ensemble, i.e., the average number of $(a,b)$ trapping sets. With
reference to Fig.~\ref{fig:RAAEncoder}, we denote by $w$ the number
of information bits that participate in an $(a,b)$ trapping set of
$\mathcal{C}^\mathrm{RAA}$. Also, let $a^i_l$, $a^o_l$, and $b_l$ be
the number of variable nodes corresponding to input bits, the
number of variable nodes corresponding to code bits, and the number
of unsatisfied checks, respectively, of code $\mathcal{C}^\mathrm{acc}_l$
involved in an $(a,b)$ trapping set of $\mathcal{C}^\mathrm{RAA}$.

To proceed, we must define the trapping set
enumerators of the component codes $A^{\mathcal{C}^\mathrm{rep}}_{w,qw}$ and
$A^{\mathcal{C}^\mathrm{acc}_l}_{a^i_l,a^o_l,b_l}$, for $l=\{1,2\}$.
Since there are no check nodes in layer 1 of the factor graph, 
$A^{\mathcal{C}^\mathrm{rep}}_{w,qw} = {K \choose w}$ is the input-output 
weight enumerator (IOWE) of
the repetition code, giving the number of codewords in $\mathcal{C}^\mathrm{rep}$
of input weight $w$ and output weight $qw$, while
$A^{\mathcal{C}^\mathrm{acc}_l}_{a^i_l,a^o_l,b_l}$ is the input-output
trapping set enumerator (IOTSE) of code $\mathcal{C}^\mathrm{acc}_l$, denoting
the number of trapping sets in $\mathcal{C}^\mathrm{acc}_l$ 
consisting of $a^i_l$ input variable nodes (i.e., variable
nodes corresponding to information bits),
$a^o_l$ output variable nodes (i.e., variable nodes 
corresponding to code bits), and $b_l$ unsatisfied checks.

With these definitions, the ensemble average TSE 
$\bar{A}^{\mathcal{C}^\mathrm{RAA}}_{a,b}$
can be computed using the uniform interleaver concept
\cite{BDMP98} as:
\vspace{-2ex}
\begin{equation}\label{eq:TSE}
    \vspace{-2ex}
\begin{split}
\bar{A}_{a,b}^{\mathcal{C}^{\mathrm{RAA}}}&=\sum_{\substack{w,a^o_1,a^o_2:~w+a^o_1+a^o_2=a\\
b_1,b_2:~b_1+b_2=b}}\frac{A^{\mathcal{C}^\mathrm{rep}}_{w,qw}A^{\mathcal{C}^\mathrm{acc}_1}_{qw,a^o_1,b_1}A^{\mathcal{C}^\mathrm{acc}_2}_{a^o_1,a^o_2,b_2}}{{N
\choose qw}{N \choose a^o_1}}\\&=\sum_{\substack{w,a^o_1,a^o_2:~w+a^o_1+a^o_2=a\\
b_1,b_2:~b_1+b_2=b}}
\bar{A}_{w,a^o_1,b_1,a^o_2,b_2}^{\mathcal{C}^{\mathrm{RAA}}},
\end{split}
\end{equation}
where $\bar{A}_{w,a^o_1,b_1,a^o_2,b_2}^{\mathcal{C}^{\mathrm{RAA}}}$
is called the ensemble-average conditional TSE.

The evaluation of \eqref{eq:TSE} requires the computation of
$A^{\mathcal{C}^\mathrm{acc}_l}_{a^i_l,a^o_l,b_l}$, which will be presented in the next section.
The extension of \eqref{eq:TSE} to more than two serially concatenated
accumulators is straightforward.

\section{Input-output trapping set enumerator for the accumulate code}
\label{sec:accenum}

In the following, we address the computation of the IOTSE
$A^{\mathcal{C}^\mathrm{acc}_l}_{a^i_l,a^o_l,b_l}$ of an accumulate code
by considering an equivalent trellis
representation of trapping sets in the factor graph.
\begin{figure}[t]
   \centering
    \includegraphics[width=6.2cm]{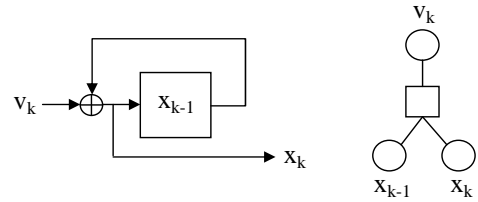}
        \vspace{-2ex}
    \caption{Block diagram and factor graph representation of an accumulate code.}
    \label{fig:Encoder}
        \vspace{-2ex}
\end{figure}
In Fig.~\ref{fig:Encoder}, the block diagram of an accumulate code
and a single section of the corresponding factor graph are depicted.
From the figure we obtain the following relation:
\begin{equation}
\begin{split}
v_k=x_{k-1}+x_k.
\end{split}
\end{equation}

\begin{figure}[t]
   \centering
    \includegraphics[width=0.98\columnwidth]{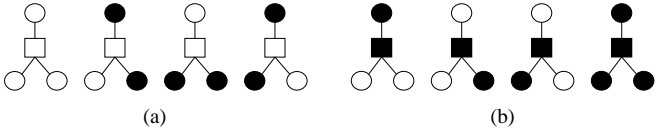}
    \vspace{-2ex}
    \caption{Factor graph representations of an accumulate code.}
    \vspace{-2ex}
    \label{fig:FactorGraph}
\end{figure}

Four different 3-tuples $(v_k,x_{k-1},x_{k})$ are possible, namely
$(0,0,0)$, $(0,1,1)$, $(1,0,1)$, and $(1,1,0)$, such that 
the parity check is satisfied. Their factor graph
representations are shown in Fig.~\ref{fig:FactorGraph}(a), where
black circles represent non-zero symbols and empty circles represent
zero symbols. Now consider an $(a,b)$ trapping set of
$\mathcal{C}^\mathrm{RAA}$, and assume that (some) of the variable
nodes of accumulate code $\mathcal{C}^\mathrm{acc}$ corresponding to
$(v_k,x_{k-1},x_{k})$ participate in the trapping set and cause an
unsatisfied check. Again, only four different configurations are
possible. They are depicted in Fig.~\ref{fig:FactorGraph}(b), 
where black circles correspond to erroneous symbols and a black
box means that the check is unsatisfied. Note that all possible trapping sets
can be obtained by properly combining the eight factor graph
sections of Fig.~\ref{fig:FactorGraph}. 

For enumeration purposes, it is simpler to refer to an equivalent 
trellis representation.
\begin{figure}[t]
  \centerline{\includegraphics[width=3.5cm]{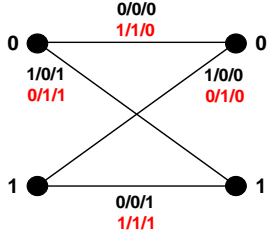}}
  \vspace{-2ex}
    \caption{Extended Trellis Section.}
    \vspace{-2ex}
    \label{fig:extendedtrellis}
\end{figure}
Assign to the variable nodes and the check nodes in
Fig.~\ref{fig:FactorGraph} that participate in a trapping set (the
black circles and boxes) the value $1$. Then the eight factor graph
sections in Fig.~\ref{fig:FactorGraph} can be conveniently
represented by the equivalent trellis section of
Fig.~\ref{fig:extendedtrellis}. We call this
the \textit{extended trellis section}
since it extends the standard trellis section of an
accumulate code to include all possible trapping sets. Each edge
between two trellis states is labeled with a binary 3-tuple
$s_i/c/s_o$, where $s_i$ denotes the input symbol, $s_o$ denotes the
output symbol, and $c$ is 1 if the check node in the corresponding
equivalent factor graph representation is unsatisfied. The four
labels in black correspond to the four configurations of
Fig.~\ref{fig:FactorGraph}(a) and define the standard trellis
section of an accumulate code, while the four labels in
red correspond to the four configurations of
Fig.~\ref{fig:FactorGraph}(b). 
Now the IOTSE of the accumulate code
can be computed from the trellis representation of
Fig.~\ref{fig:extendedtrellis} by considering a trellis consisting
of $N$ concatenated trellis sections like the one in
Fig.~\ref{fig:extendedtrellis} and enumerating all possible paths.
The IOTSE is given in closed form in the following Theorem.

\begin{thm}\label{thm:iotse}
Let $(a^i,a^o,b)$ be a trapping set with $a^i$ information variable
nodes, $a^o$ code variable nodes, and $b$ unsatisfied checks. The
\textit{input-output trapping set enumerator} (IOTSE) for the
rate-1, memory-1, convolutional encoder $\mathcal{C}^\mathrm{acc}$ with
generator polynomial $g(D) = 1/(1 + D)$, terminated to the all-zero
state at the end of the trellis, and with input and output block
length $N$, can be given in closed form as:
\begin{equation}\label{eq:TrappingWEF}
\begin{split}
A_{a^i,a^o,b}&=\sum_{m}\sum_{n}{N-a^o
\choose m}{a^o-1 \choose m-1}\cdot\\
&\cdot{a^o-m \choose \frac{a^i+b}{2}-n-m}{N-a^o-m \choose n}{2m
\choose \frac{a^i-b}{2}+m}, \end{split}
\end{equation}
where $m$ and $n$ must satisfy the constraints
\begin{equation}
\begin{split}
m&\geq \frac{|a^i-b|}{2},~~~m\leq\min\{a^o,N-a^o\},\\
n&\geq \frac{a^i+b}{2}-a^o,~~~n\leq N-a^o-m.
\end{split}
\end{equation}
\end{thm}
\begin{proof}
Consider the extended trellis section of the encoder $g(D)=1/(1+D)$
in Fig.~\ref{fig:extendedtrellis}. Denote by $n$ the number of
length-one error events $1/1/0$ from the zero state to the zero state,
called type-1 error events, and by
$m$ the number of error events that leave the zero state, and remerge
later to the zero state, called type-2 error events. Further, let
$a^i$, $a^o$, and $b$ be the number of information variable nodes, the
number of code variable nodes, and the number of unsatisfied checks,
respectively, participating in the trapping set. Also, let $w^t$
denote the total input weight associated with the transitions
$0\rightarrow1$ (from state zero to state one) and $1\rightarrow0$
(from state one to state zero) in the $m$ type-2 error events.

Only type-2 error events are responsible for the weight at the
output of the accumulator. From \cite{div98}, we know that the
number of permutations of $m$ type-2 error events resulting in an output
weight of $a^o$ is 
\begin{displaymath}
\binom{N-a^o}{m} \binom{a^o-1}{m-1}.
\end{displaymath}
(Here, the transitions away from and back to the zero state are not only caused by
the input weight $w^t$ but also by $2m-w^t$ unsatisfied check nodes.)
The $m$ type-2 error events include
$a^o-m$ transitions from the one state to the one state
($1\rightarrow1$), and the input weight associated with the
transitions $1\rightarrow1$ is 
\begin{equation}\label{eq:w11}
w_{1\rightarrow 1}=a^i-n-w^t.
\end{equation}
Moreover, the following equality holds:
\begin{equation}\label{eq:a}
w^t=\frac{a^i-b}{2}+m.
\end{equation}
Also, due to termination, $a^i+b$ is even. 
\vspace{0.4ex}From \eqref{eq:w11} and \eqref{eq:a} it now follows that
$w_{1\rightarrow 1}=\frac{a^i+b}{2}-n-m$. This weight can be ordered in 
\vspace{0.4ex}${a^o-m\choose \frac{a^i+b}{2}-n-m}$ different ways,
which gives the third binomial coefficient in (\ref{eq:TrappingWEF}).
On the other hand, there are $N-a^o-m$ transitions from the zero state 
to the zero state with an associated input weight $n$. 
\vspace{0.5ex} Therefore, we obtain the term 
\vspace{0.5ex}${N-a^o-m\choose n}$.
The last binomial coefficient in (\ref{eq:TrappingWEF}) results from
the ordering of the $w^t$ ones in the 
\vspace{0.4ex}$2m$ transitions $0\rightarrow1$ and $1\rightarrow0$, in 
\vspace{0.4ex}${2m\choose \frac{a^i-b}{2}+m}$ ways.

To summarize, the number of paths in the extended trellis consisting of
$n$ type-1 error events and $m$ type-2 error events is given by:
\begin{small}
\begin{displaymath}
\!\! {N-a^o\choose m} \!\! {a^o-1 \choose m-1} \!\! {a^o-m \choose
\frac{a^i+b}{2}-n-m} \!\! {N-a^o-m \choose n} \!\! 
{2m \choose \frac{a^i-b}{2}+m}\!.
\end{displaymath}
\end{small}

The result for the encoder $g(D)=1/(1+D)$ follows by summing over
all possible values of $n$ and $m$.
\end{proof}
\begin{cor}
For $b=0$, the expression in (\ref{eq:TrappingWEF}) reduces to the 
well-known IOWE for the rate-1, memory-1, accumulate code \cite{div98}.
\end{cor}

\section{Asymptotic ensemble trapping set enumerator}

In order to determine the asymptotic spectral shape of the trapping sets
associated with a particular code ensemble, 
as the block length $N$ tends to infinity,
we define the normalized logarithmic
asymptotic TSE $r^\mathcal{C}(\alpha,\beta)$ of a code 
ensemble $\mathcal{C}$ as
\vspace{-1ex}
\begin{equation}\label{eq:spectralshape}
\vspace{-1ex}
r^{\mathcal{C}}(\alpha,\beta) = \limsup_{N\rightarrow
\infty}\frac{\ln \bar{A}_{a,b}^{\mathcal{C}}}{N},
\end{equation}
where $\alpha= a/N$, $\beta=b/N$, and the supremum is taken over all
intermediate variables. We also define the functions
$f^{\mathcal{C}^\mathrm{rep}}$ and $f^{\mathcal{C}^\mathrm{acc}_l}$ as the asymptotic
behavior of the IOWE of a repeat code and the asymptotic behavior of
the IOTSE of an accumulate code $\mathcal{C}^\mathrm{acc}_l$, respectively:
\vspace{-1ex}
\begin{equation}\label{eq:functionsf}
\vspace{-1ex}
\begin{split}
f^{\mathcal{C}^\mathrm{rep}}(\omega)&=\lim_{N\rightarrow
\infty} \frac{\ln A^{\mathcal{C}^\mathrm{rep}}_{w,qw}}{N}\\
f^{\mathcal{C}^\mathrm{acc}_l}(\alpha^i_l,\alpha^o_l,\beta_l)&=\lim_{N\rightarrow
\infty}\frac{\ln A^{\mathcal{C}^\mathrm{acc}}_{a_l^i,a_l^o,b_l}}{N}, \quad l=1,2,
\end{split}
\end{equation}
where $\omega=w/K$, $\alpha^i_l= a_l^i/N$, $\alpha^o_l= a_l^o/N$, and
$\beta_l= b_l/N$.

Using Stirling's approximation for binomial coefficients ${n \choose
k}\stackrel{n \to
\infty}{\longrightarrow}e^{n\mathbb{H}\left(\frac{k}{n}\right)}$,
where $\mathbb{H}(\cdot)$ is the binary entropy function with
natural logarithms, the functions in (\ref{eq:functionsf}) can be
written as:
\begin{equation}\label{eq:functionfrep}
f^{\mathcal{C}^\mathrm{rep}}(\omega) = \frac{1}{q}\mathbb{H}(\omega),
\end{equation}
and
\vspace{-1ex}
\begin{equation}\label{eq:functionfacc}
\vspace{-1ex}
\begin{split}
&f^{\mathcal{C}^\mathrm{acc}_l}(\alpha^i_l,\alpha^o_l,\beta_l)= \sup_{\mu_l,\nu_l}~~(1-\alpha^o_l)\mathbb{H}\left(\frac{\mu_l}{1-\alpha^o_l}\right)+\\
&+\alpha^o_l\mathbb{H}\left(\frac{\mu_l}{\alpha^o_l}\right)+(\alpha^o_l-\mu_l)
\mathbb{H}\left(\frac{\alpha^i_l+\beta_l-2(\nu_l+\mu_l)}{2(\alpha^o_l-\mu_l)}\right)+\\
&+(1-\alpha^o_l-\mu_l)\mathbb{H}\left(\frac{\nu_l}{1-\alpha^o_l-\mu_l}\right)+2\mu_l
\mathbb{H}\left(\frac{\alpha^i_l-\beta_l+2\mu_l}{4\mu_l}\right),
\end{split}
\end{equation}
where we have defined the normalized quantities $\mu_l= m_l/N$ and
$\nu_l= n_l/N_l$.

Then, using (\ref{eq:functionsf}-\ref{eq:functionfacc}) and
(\ref{eq:TSE}) in (\ref{eq:spectralshape}), the asymptotic TSE of a
code ensemble $\mathcal{C}^{\mathrm{RAA}}$ can be written as:
\begin{equation}\label{eq:asymptTSE}
\begin{split}
r^{\mathcal{C}^{\mathrm{RAA}}}(\alpha,\beta)
&=\sup_{\stackrel{\alpha=\omega/q+\alpha^o_1+\alpha^o_2}{\beta=\beta_1+\beta_2}}
f^{\mathcal{C}^\mathrm{rep}}(\omega)+f^{\mathcal{C}^\mathrm{acc}_1}(\omega,\alpha^o_1,\beta_1)+\\
&+f^{\mathcal{C}^\mathrm{acc}_2}(\alpha^o_1,\alpha^o_2,\beta_2)-\mathbb{H}(\omega)-\mathbb{H}(\alpha^o_1),
\end{split}
\vspace{-2ex}
\end{equation}
with the constraints $\alpha=\frac{\omega}{q}+\alpha^o_1+\alpha^o_2$
and $\beta=\beta_1+\beta_2$.

\section{Numerical Evaluation}

\begin{figure}[t]
   \centering
    \includegraphics[width=0.87\columnwidth]{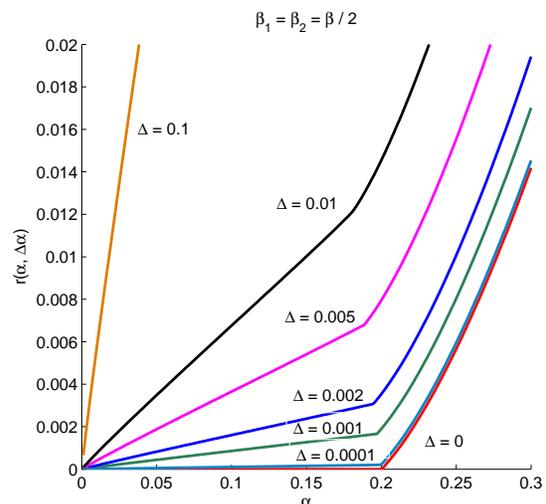}
    \vspace{-4ex}
    \caption{Asymptotic TSE for different values of $\Delta$ and $\beta_1=\beta_2=\beta/2$.}
    \vspace{-3ex}
    \label{fig:delta05}
\end{figure}

In this section, we present a numerical evaluation of \eqref{eq:asymptTSE}. 
Following \cite{AsRD07}, in the curves for the asymptotic TSE that we present,
we keep the ratio $\Delta=\beta/\alpha$ of 
unsatisfied check nodes to erroneous variable nodes constant and 
compute $r(\alpha, \Delta\alpha)$ for varying values of $\alpha$. 
In Fig.~\ref{fig:delta05}, the unsatisfied checks in the RAA code ensemble
are equally distributed between the middle and inner accumulator, 
i.e., $\beta_1=\beta_2=\beta/2$.
For $\Delta=0$, when no unsatisfied checks are present in the factor graph, 
the spectral shape $r(\alpha,0)$ exhibits a zero stretch in the beginning 
and turns positive when the number of codewords with normalized 
weight $\alpha$ starts to grow exponentially in $N$ with increasing $\alpha$. 
The presence of unsatisfied checks in the factor graph results in a 
positive initial slope, and we observe a quasi-linear increasing first section
of the curve, until there is a discontinuity in the slope.
In the second section, the slope of the curve  is similar for all values of
$\Delta$, and the curve shifts to the left with increasing $\Delta$. 
Also, as the fraction of unsatisfied checks $\Delta$ increases, 
the slope in the first section also increases.
\begin{figure}[t]
   \centering
    \includegraphics[width=0.87\columnwidth]{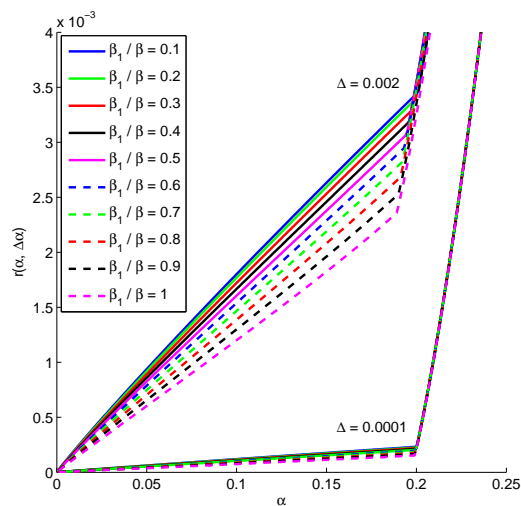}
    \vspace{-4ex}
    \caption{Asymptotic TSE for different fractions $\beta_1/\beta$.}
    \vspace{-3ex}
    \label{fig:varBeta}
\end{figure}
Because of the large number of parameters involved in taking the supremum in 
\eqref{eq:asymptTSE}, it is difficult to draw general conclusions about the
trapping set structures that are most likely to cause decoding failures.
The structure of a trapping set is greatly influenced by the choice of
these parameters. We are primarily concerned with trapping set
configurations that lead to decoding errors and this requires $\omega>0$.
The choice of the parameters $\beta_1$ and $\beta_2$ determines how many
unsatisfied checks are associated with the middle and inner accumulator,
respectively. For instance, in the extreme case of $\beta_1=\beta$ 
and $\beta_2=0$, all the unsatisfied checks are associated with the middle
accumulator, and there are no unsatisfied checks in the graph of the inner
accumulator.

In Fig.~\ref{fig:varBeta} we vary the ratio $\beta_1/\beta$, the 
fraction of unsatisfied check nodes associated with the middle accumulator 
in the RAA code ensemble. For larger $\beta_1/\beta$, when relatively 
more unsatisfied check nodes are present in the
middle accumulator, the slope in the first section is smaller and 
the influence of varying $\beta_1/\beta$ on the slope becomes greater as
$\beta_1/\beta$ gets closer to one.
However, the influence that varying $\Delta$ has on the slope is much greater 
than the influence of varying $\beta_1/\beta$. Also, for larger values of $\Delta$, 
the variance of the curves with $\beta_1/\beta$ is greater.

\begin{figure}[t]
   \centering
    \includegraphics[width=0.88\columnwidth]{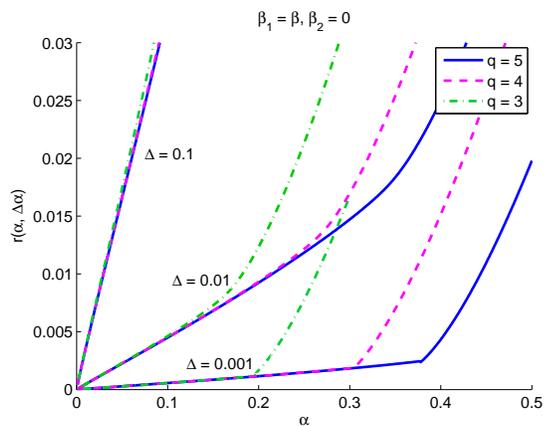}
    \vspace{-3ex}
    \caption{Asymptotic TSE of the RAA code ensemble for different repetition 
    	factors $q$.}
    \vspace{-1ex}
    \label{fig:varQ}
\end{figure}
\begin{figure}[t]
   \centering
    \includegraphics[width=0.88\columnwidth]{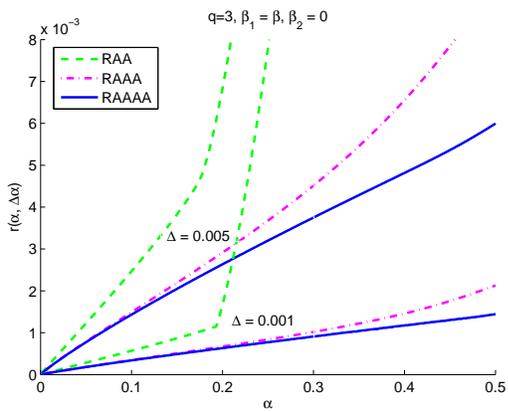}
    \vspace{-3ex}
    \caption{Asymptotic TSE for the RAA, RAAA, and RAAAA code ensembles.}
    \label{fig:varL}
    \vspace{-3ex}
\end{figure}
In Figs.~\ref{fig:varQ} and \ref{fig:varL} we display the influence 
of the repetition factor $q$ and the number of concatenated
accumulators, respectively, on the shape of the asymptotic TSE.
In both cases the greatest effect on the slope in 
the first section was observed when $\beta_1=\beta$, i.e., when all
unsatisfied check nodes are associated with the outermost accumulator. 
In the other extreme case, when all the unsatisfied 
check nodes are associated with the inner accumulator,
the slope in the first section did not change.
In Fig.~\ref{fig:varQ}, the reduction in slope in the first section
caused by increasing the repetition factor $q$ is only marginal 
and the curves almost lie on top of each other.
However, increasing the number of serially concatenated accumulators
decreases the slope in the first section, as can be seen in Fig.~\ref{fig:varL}.
Finally, we note that increasing the repetition factor $q$ or adding more
accumulators increases the minimum distance of the code, and thus
the transition from the quasi-linear section of the asymptotic 
TSE to the more steeply increasing section takes place at 
higher values of $\alpha$.

\section{Conclusions}

We have presented a simple closed form method to enumerate general $(a,b)$ 
trapping sets for RAA code ensembles. The trapping set enumerator is first 
obtained for finite block lengths $N$ and its asymptotic expression is derived
by letting $N$ go to infinity. Similar to \cite{Mil07} and \cite{AsRD07}, 
we observe that, when unsatisfied check nodes are present in the factor graph,
the asymptotic TSE lies strictly above the asymptotic spectral shape 
for the case when no unsatisfied check nodes exist in the graph.
Although the RAA code ensemble is asymptotically good and exhibits minimum
distance growing linearly with block length, in contrast to regular and some 
protograph-based irregular LDPC codes, there exists no region where the 
minimum trapping set size grows linearly with block length. It can, at best,
grow only sublinearly in the block length, since the asymptotic TSE of the 
RAA code ensemble is always positive if unsatisfied check nodes are 
present in the graph.
While the method presented in this paper allows us to enumerate all 
general $(a,b)$ trapping sets, the influence that 
these trapping sets have on the error floor must still be evaluated separately. 
As noted earlier, the probability that the decoder gets stuck in particular types of
trapping sets depends on the channel, the decoding algorithm, and the
particular decoder implementation. 
In future work we hope to evaluate this probability for the turbo decoder and
the belief propagation decoder, in order to obtain a reliable estimate of the height
of the error floor for RMA codes.

\bibliographystyle{ieeetr}
{\small \itemsep 10ex
\bibliography{biblio}}

\end{document}